\documentstyle[prl,multicol,aps]{revtex}

\def\ll{\label} \def\re{\ref} \def\c{\cite}  
\def\r1{(\ref{$1})}   
 \def\kap{\kappa}

  \def\ep{\epsilon} \def\th{\theta}
\def\ba{\begin{array}{c}}  
\def\ea{\end{array}}   
  \def\de{\delta} 
\def\ov{\over}   \def\l{\left}
\def\l({\left(} \def\r){\right)} \def\r{\right} \def\rw{\rightarrow}
   
\def\be{\begin{equation}} \def\bc{\begin{center}} \def\ec{\end{center}}
\def\bit{\begin{itemize}} \def\eit{\end{itemize}} \def\ee{\end{equation}}
\def\ed{\end{document}} \def\bea{\begin{eqnarray}} \def\eea{\end{eqnarray}}
\def\efr{\end{flushright}}  
\begin{document} \title{
Exact solution of double-$\delta$ function Bose gas through interacting
anyon gas}

\author{
Anjan Kundu \footnote {email: anjan@tnp.saha.ernet.in} 
 \footnote {Address during May-July, 1999:\\
           Inst. fur Theoretische Physik,
           Appelstr. 2,
           Universitat Hannover
           D-30167 Hannover, GERMANY}  \\  
  Saha Institute of Nuclear Physics,  
 Theory Group \\
 1/AF Bidhan Nagar, Calcutta 700 064, India.
 }
\maketitle
%
%\vskip 1 cm
%\rightline{hep-th/9811247, SINP/TNP/98-27}

%Running title: {\it double-delta bose gas}
\begin{abstract} 
%------------------------------------------------------------
 1d Bose gas interacting through $\delta, {\de}^{'} $ and double-$\delta$
 function potentials is shown to be equivalent to a $\delta$ anyon gas
allowing exact
 Bethe ansatz solution. In the { noninteracting} limit it describes an ideal
gas with generalized exclusion statistics and solves some recent
controversies.
 %45words-----------------------------------------
\end{abstract}

%\medskip
 PACS numbers: 
05.30.Jp,
03.70.+k 
11.55.Ds, 
71.10.Pm,

The concept of particles with generalized exclusion statistics (GES)
introduced by Haldane \c{ges} has important consequences \c{wuyu} in
describing
 1d non Fermi-liquids \c{luttinger}, which in turn is believed to be related
\c{wen} to the edge excitations in fractional quantum Hall effect \c{fqhe}.
 On the other hand, inspired by the success of the Chern-Simon theory, an
 attempt was made recently \c{rabello} to describe a 1d ideal gas with GES
in the framework of a gauge field model. However, in a subsequent paper
\c{jakiws} the previous result was shown to be wrong and some other
 conclusion was offered.  Our aim here is to deal primarily with a 1d Bose
 gas interacting through  double-$\delta$ function potentials together with
the well known
$\delta$    and derivative $\delta$-function interactions.
We show  that this interacting model with several singular potentials is
equivalent to a 1d gas with GES (which we call anyon for brevity)
interacting via $\delta$-function potential only. This $\delta$-anyon gas is
found to be exactly solvable by the coordinate Bethe ansatz (CBA) just like
its bosonic counterpart, contradicting the common belief \c{gutkin} that the
CBA is applicable only to models with symmetric or antisymmetric wave
functions. Remarkably, at the limit of vanishing
 interaction the anyon gas becomes free and gauge  equivalent 
to a  related  model
proposed in \c{rabello}. This shows that, though the explicit wave
function and  the $N$-body Hamiltonian of \c{rabello} are
 not exact, the conclusion it arrived at
is basically correct. Therefore, while the  error in the treatment of 
\c{rabello} was  detected in \c{jakiws}, the
source of this error and the possible way to rectify it becomes evident from
our result.

We start with a 1d  Bose gas interacting through generalized
$\delta$ function potentials as 
  \be {H}_N
=-\sum_k^N \partial^2_{x_k}+  \sum_{<k,l>} 
\de ({x_k-x_l})\left( c+i \kap 
 (\partial_{x_k}+\partial_{x_l})\right) + \gamma_1 \sum_{<j,k,l>}
 \de ({x_j-x_k})
 \de ({x_l-x_k}) + \gamma_2 \sum_{<k,l>}
 (\de  ({x_k-x_l}))^2.  
\ll{hdd} \ee
This model  was briefly considered and readily
discarded in \c{dipti} as  too difficult a problem to solve. Notice however
that for $\gamma_a=0, a=1,2,$ i.e. without the double-$\delta$ potentials,
 it has
various
 exactly  solvable limits. For example, for $\kap=0, c \neq 0$
  the model becomes the well known $\delta$-Bose gas \c{delta},
 while for $\kap \neq 0, c = 0$ it corresponds to Bose gas with 
  ${\delta } ^{'}$ interaction \c{snirman}. Both these cases are not only
 exactly solvable by CBA, but also  represent quantum integrable systems
allowing $R$-matrix solution. This can be
proved through their connection with the quantum integrable nonlinear
Schr\"odinger equation (NLSE) \c{qnls} and  derivative NLSE
\c{qdnls}, respectively.
  Even the mixed case with $\kap \neq 0, c \neq 0$ is solvable through CBA
\c{gutkin,snirman}, though as a quantum model it does not allow a $R$-matrix
solution. Nevertheless for $\gamma_a \neq 0$, i.e. with the inclusion of
highly singular double-$\delta$ function interactions, the solvability of the
model is completely lost and the application of the CBA becomes
problematic due to the presence of tree-body interacting terms. We ask
therefore,  whether for some choice of the coupling constants
$\gamma_a$ other than zero, this difficulty could still be avoided  and  
 the solvability of the model be restored.  We find the answer to be
 affirmative and in particular, for $\gamma_a =\kap^2$ the model becomes
 equivalent to a
$\delta$ function anyon gas, which appears to be exactly solvable similar to
the well known bosonic case obtained  at $\gamma_a=\kap=0$.

Instead of attacking  model (\re{hdd}) directly, our strategy would be to
transform it into some equivalent tractable problem. For this we notice
that, parallel to the relation between the $\de $ Bose gas and the NLS model
 \c{gutkin2},  the generalized interacting bosonic model (\re{hdd}) can be
considered to be the $N$-particle Hamiltonian of
the nonlinear field model 
  \be {H}
=\int dx \left[ : \left( (\psi^\dag_x \psi_x  + c  \rho^2
+i \kap  \rho (\psi^\dag \psi_x- \psi^\dag_x \psi) \right):
 +\kap^2
(\psi^\dag \rho ^2 \psi) \right], \ \ \ \ \rho \equiv (\psi^\dag \psi)
\ll{hke} \ee
 involving bosonic   operators
 :$ [\psi (x), \psi^\dag(y)]= \de(x-y)$. In  (\re{hke})
 we have chosen $\gamma_1=\gamma_2=\kap^2  $  and introduced 
notation $\ : \ \ : \ $ to indicate normal ordering (NO) in bosonic
operators. Restricting  now to the $|N>$ particle
state and defining the $N$-particle wave function as
\be 
\Phi (x_{i_1},x_{i_2}, \ldots ,x_{i_N})= <0|\psi(x_{i_1})\psi(x_{i_2})
\cdots \psi(x_{i_N}) |N>,
\ll{wavef}\ee
we can generate all  terms of (\re{hdd}) starting from  
(\re{hke}). For example, the last term  in (\re{hke}):
 $\ \int dx (\psi^\dag \rho ^2 \psi) \ $ is equivalent to  two normally
ordered  terms like
 $ \int dx : ( \rho^3 + \rho^2 \int dy (\de (x-y))^2):\ \ $. The first one
when  acts 
on the state $|N>$, its three $\psi (x)$ operators in passing through the
creation operators at points $x_j, x_k, x_l$  in $|N>$
 would produce sum of  terms with product of three
$\de$ functions having arguments $(x-x_j),(x- x_k)$ and $(x-x_l).$ On
integration by $x$ they would generate     the double-$\de$ function
potential $ 
 \de ({x_j-x_k})
 \de ({x_l-x_k}).$
 Note that this is a 3-body term and
would not contribute  in  2-body bosonic interactions. On the other hand,
 the
second term acting  
on  $|N>$, would give rise to the sum of  terms like  $ 
 \de ({x_k-x_l})
 \de ({x_k-x_k}) \approx 
 (\de ({x_k-x_l}))^2 .$ Similarly
other terms with $\de ^{'}$ and
$\de$ function interactions are   obtained in (\re{hdd})  from
  (\re{hke}).
   
Our next step is to define a gauge transformed operator
\be
\tilde \psi (x)= e^{-i \kap  \int^x_{- \infty}
  \psi^\dag (x') \psi (x') dx'} \psi (x) 
   \ll{tpsi} \ee
along with its conjugate $\tilde \psi^{ \dag} (x)$. We may check that 
the derivatives and products of the transformed operators
are related to the old ones in   the following way.
\bea
 \vdots \ (\tilde \psi^\dag \tilde \psi)^2 \ \vdots &= &: (\psi^\dag \psi)^2:
\\ 
 \vdots  \ \tilde \psi^\dag_x \tilde \psi_x \  \vdots 
&=& (\psi^\dag_x +i \kap  \psi^\dag \rho) 
(\psi_x -i \kap  \rho \psi) = :(\psi^\dag_x \psi_x 
+i \kap  \rho (\psi^\dag \psi_x- \psi^\dag_x \psi)): +\kap^2
(\psi^\dag  \rho ^2 \psi)    ,
\ll{tpsipsi}\eea
where $ \ \vdots \ \  \vdots \ $ stands for NO with respect to the
transformed operator (\re{tpsi}), which not necessarily coincides with 
the bosonic NO as evident from (\re{tpsipsi}).  
Using these relations  therefore one rewrites  Hamiltonian
(\re{hke}) in the form 
\be
\tilde H=\int dx  \vdots \left(
\tilde \psi^\dag_x \tilde \psi_x 
  + c (\tilde \psi^\dag \tilde \psi)^2 \right) \vdots
\ll{thnls}\ee
Note however that  inspite of the same form as NLSE, 
 (\re{thnls}) is not same as the known model, since  the fields involved are
no longer bosonic operators but exhibit {\it anyon} like properties
\be
\tilde \psi^\dag (x_1) \tilde \psi^\dag (x_2)=e^{i \kap \ep (x_1-x_2)}
\tilde \psi^\dag (x_2)\tilde \psi^\dag (x_1) , 
\ \tilde \psi (x_1) \tilde \psi^\dag (x_2)=e^{-i \kap \ep (x_1-x_2)}
\tilde \psi^\dag (x_2)\tilde \psi (x_1)+ \de (x_1-x_2)\ll{cranyon}\ee
etc. where
%1
 \be \ep (x-y)= \pm 1 \ \mbox{for}\ x >y, \ x< y \  
\mbox{and} 
\ =0  \ 
\mbox{for} \ x = y, \ll{e} \ee
which may  be expressed also through the symmetrical unit-step function
\c{handbook}. 
This means that 
   the bosonic commutation relation (CR)
$ [\tilde \psi (x), \tilde \psi^\dag(y)]= \de(x-y)$ remains valid 
at the coinciding points.
%1
 These relations
 can be checked easily by using realization (\re{tpsi}) through bosonic
fields.

 For finding    $N$-body
Hamiltonian corresponding to (\re{thnls}), we observe that 
operator $\tilde \psi
(x)$  in passing through the string of anyonic creation operators in $|
\tilde N>$
would  pick up first a phase $e^{- i \kap \sum_{i <k} \ep (x-x_i)}$ due to
 (\re{cranyon}) and then leave a $\de (x-x_k)$ at $x_k$ due to its standard CR
at the coinciding points. The phase factor however is canceled subsequently
 when
the associated $\tilde \psi^{\dag}
(x)$ also passes  through the same creation
operators and comes to the point $x_k$. This happens 
 due to the opposite signs of the phases as seen from  (\re{cranyon}). 
Therefore finally,  similar to  the bosonic model
 one obtains  a $\de$ function
interacting gas
  \be \tilde {H}_N
=-\sum_k^N \partial^2_{x_k}+  c\sum_{<k,l>} \de ({x_k-x_l})
\ll{had}\ee
However, in contrast to the standard case the wave function now 
exhibits a generalized
symmetry
\be 
\tilde \Phi (x_{1}, \ldots ,x_{i}, \ldots, x_{j}, \ldots,x_{N})
= e^{-i \kap \left( \sum_{k=i+1}^ j 
\ep (x_{i} - x_{k})- \sum_{k=i+1}^ {j-1} 
\ep (x_{j} - x_{k}) \right)} 
\tilde \Phi (x_{1}, \ldots ,x_{j}, \ldots, x_{i}, \ldots,x_{N})
\ll{waves}\ee
dictated by the operator relations (\re{cranyon}) and the model is defined 
 in an infinite-interval space $R$.
%2
It should be emphasized that due to the validity of  bosonic CR 
at  coinciding points in the
anyonic relations (\re{cranyon}), in  the induced wave
function (\re{waves}) the phase  factor with 
${ \ep (x_{l}-x_{k})}$  vanishes  at $x_{l}=x_{k} $ 
making it  well defined.
% since $ \ep (x_{l}-x_{k}){\mid_{0}}= 0. $
%2
Note  that the commutation relations (\re{cranyon}) and the
symmetry of the wave function  (\re{waves}) for this 1d model
 are remarkably consistent
with the genuine anyonic behaviors in 2d \c{partha}. Like the
anyonic property, while $\tilde \psi^\dag$ is $q^{-1}=
e^{i \kap \ep (x_{1} - x_{2})}$-symmetric, the wave function $
\tilde \Phi (x_{1}, x_{2})$ is $q$-symmetric, so that the $|\tilde 2>$-particle
 state remains invariant under permutation of coordinates.
%3
It is  evident that the range of $\kappa$ is sufficient to be 
restricted in the interval of
$2 \pi$ by choosing $- \pi \geq \kappa \leq  \pi $ and   
any nonzero value of $\kappa$
 affects the symmetry
under reflection of coordinates, close to the 2d anyon case. 
%3 
Another parallel of multi-anyon wave function \c{partha} can be observed in
(\re{waves}), i.e. the phase factor appearing under exchange of two of its
arguments also depends
  on the intermediate coordinates.      

We have converted thus the original 
eigenvalue problem related to (\re{hdd})
to an equivalent  one: $\tilde H_N  \tilde \Phi
(x_{1},  \ldots, x_{N}) = E_N \tilde \Phi (x_{1},  \ldots, x_{N})$ for
 (\re{had}) acting on anyon-type wave function (\re{waves}). Since the Bethe
ansatz solution is meant apparently for the (anti-)symmetric wave functions
only \c{gutkin}, the present problem claims novelty.
 However, we find that
 the CBA is applicable here  with equal success, if one modifies the Bethe
 ansatz for the wave function appropriately:
\be 
\tilde \Phi (x_{1}, \ldots,x_{N})=\Phi_A (x_{1}, \ldots,x_{N}) \Phi_B
 (x_{1}, \ldots,x_{N}).
\ll{BAany}\ee
Here $\Phi_B$ is the symmetric function in the standard Bethe ansatz form
\c{delta} 
\be
 \Phi_B
 (x_{1}, \ldots,x_{N})=\sum_P A(P) 
e^{i \sum_{j} 
x_{j} k_{P_j}}
\ll{BA}\ee
defined in the primary region $R_1:
 \ \ x_{1}\leq x_{2}\leq \ldots\leq x_{N}
\ \ ,$  while  $ \Phi_A $ is the additional anyonic part 
given as 
\be
\Phi_A (x_{1}, \ldots,x_{N})= e^{-i {\kap  \ov 2} \Lambda 
(x_{1}, \ldots,x_{N})}, \ \ \mbox {with} \ \ \Lambda \equiv {\sum_{i<j} 
\ep (x_{i} - x_{j})}, \ll{wany}\ee
with $
\ep (x_{i} - x_{j}) $ as defined in (\ref{e}).
Remarkably,  the discontinuity in the derivative of the
wave function (\re{BAany}) at the boundary
\be
(\partial_{x_l}- \partial_{x_{k}})\tilde \Phi {\mid_+}-
(\partial_{x_l}- \partial_{x_{k}})\tilde \Phi {\mid_-} = {c}
 \tilde \Phi {\mid_{0}}
\ll{dpsi}\ee
with the notation ${\mid_{\pm}}=\mid_{x_l=x_{k}^{\pm}} $ and 
${\mid_{0}}= {\mid_{x_{l} = x_{k}}},$
 determines the scattering amplitude in the same way as in    case
of $\delta$-Bose gas. This becomes possible since using (\re{BAany})
, (\re{wany}) 
%4
along with  (\re{e})
%4
 one obtains
\bea
\tilde \Phi (x_{1}, \ldots,x_{N})\mid_{+}&=& e^{-i {\kap  \ov 2}(-1+S)} \Phi_B
 (x_{1}, \ldots,x_{N})\mid_{x_l=x_{k}^{+}},  \nonumber \\ 
\tilde \Phi (x_{1}, \ldots,x_{N})\mid_{-}&=& e^{-i {\kap  \ov 2}(+1+S)} \Phi_B
 (x_{1}, \ldots,x_{N})\mid_{x_l=x_{k}^{-}},  \nonumber \\ 
\tilde \Phi (x_{1}, \ldots,x_{N})\mid_{0}&=&
 e^{-i {\kap  \ov 2}(S)} \Phi_B
 (x_{1}, \ldots,x_{N})\mid_{x_l=x_{k}} 
 \ll{wany3}\eea
where $S=\Lambda -\ep (x_{k} - x_{l})$.
%5
Contributions coming from the derivatives of other
$\epsilon (x_i-x_j)$ factors (as $\de$ functions) in (\re{dpsi})
 clearly cancel  each
other, transforming it
 consequently  to an equation only for the symmetric part
  of the wave
function
 in the standard form
\be
(\partial_{x_l}- \partial_{x_{k}}) \Phi_B (x_{1}, 
\ldots,x_{N}) {\mid_+}= 
{\tilde c}
  \Phi_B(x_{1}, \ldots,x_{N}) {\mid_{0}},
\ll{dpsib}\ee
but with  modified coupling constant $\tilde c={c \ov 4 \cos {\kap \ov
2}}$.
Note that the
 singularity of the original bosonic problem is reflected in the
discontinuity of the anyonic wave function (\ref{BAany}). Such discontinuity
at the boundaries of different regions, though somewhat unusual in CBA, has
 been observed  recently in other context \c{pla98}. In the present case,
 this also does not
affect the physical picture, as evident from the
reduction of the problem to (\ref{dpsib}) for continuous wave
functions.
%5
 Therefore following  arguments of the bosonic model \c{delta}, one
can reduce (\re{dpsib}) further to the region $R_1$ involving only 
adjacent $k'$s, i.e. with ${x_l} \rw {x_{k+1}} $ and using  the Bethe
ansatz (\re{BA}) calculate the $2$-particle scattering amplitude as 
$e^{i \th_{l l+1} }= {k_l-k_{l+1}-i {\tilde c \ov 2} \ov 
k_l-k_{l+1} +i {\tilde c \ov 2}}=e^{-i \th_{l+1 l} }.$ 
Mark however, that in
contrast to \c{delta} the coupling constant in this anyonic case has been
changed. At
$\kap =0$ one recovers the bosonic case, while $\kap= \pm \pi$ gives hard core
repulsion $c \rw \infty ,$ simulating  fermionic model. In the present
infinite-interval space the values of $\{k\}$ have no restriction. 
However if we restrict to the interval $0 \leq x_j \leq L ,$ the boundary
condition would be twisted like
$ \  \ 
\tilde \Phi (x_{1+L},\ldots,x_{N})
=  e^{i {\kap } 
(N-1)} \tilde \Phi (x_{1}, \ldots,x_{N}) \ \ $ and one would obtain the 
determining equations for $k'$s  as
$ \ \  
k_{j}= -{1 \ov L} \sum_{s=1}^N \tilde \th_{js} +{2 \pi \ov L} n_j
 + {\kap  \ov L} (N+1-2j), \ \ j=1, \ldots ,N   
\ \ $ with  $n_j$   an integer.  Here we have redefined 
$ \tilde \th_{js}= \th_{js} - \kap \ep (j-s)
 = -\tilde \th_{sj}$  to introduce 
an anyonic scattering phase,
since such a particle in passing through  others would  pick
 up a phase $e^{\pm i
\kap}$ depending on its relative position.
 The energy eigenvalue of the system
 $E_N= \sum_{j}^N k_j^2, $
though has the  same form  as in bosonic model,  acquires in  effect
different value due to changed coupling constant.
%6
We may mention here that the possibility of gauging away certain
multi-species
 fermionic interaction by introducing a twisting in the boundary condition
has been  observed recently  \c{ss-prl98}.
Similar spirit of the present result in a totally
 different context  might therefore be an indication of
a  deeper generality. 
%6

Now at $c \rw 0$ limit of (\re{had}) we recover the results related to
a GES ideal gas having  properties like (\re{cranyon}) and (\re{waves}).
As we have shown, this {free} anyonic Hamiltonian  would be
 equivalent to the $N$-bosonic model (\re{hdd}) at $c=0,\gamma_a=\kap^2.$
Establishing such a fact through direct gauge transformation of the wave
function was the aim of \c{rabello}.
   However this seems to be difficult to achieve, in which
investigations      \c{rabello,jakiws}    were concentrated.
 We have avoided this difficulty by showing the equivalence
through the gauge relation between the related field models (\re{thnls}) and
(\re{hke}).   Mark also that   the  field 
 model (Eqn. (8)) of \c{rabello} differs from that  (Eqn. (6)) of \c{jakiws}
and ours by a crucial term.
  We note again that (\re{hdd}) even with $c=0$ involves 3-body
terms, ignoring which naturally would  make it not equivalent to the free
anyonic Hamiltonian. In both \c{rabello} and \c{jakiws}  the authors
 worked with a 2-body bosonic Hamiltonian which could not give the right
equivalence.
      
 The author likes to thank  Partha Mitra for
helpful discussions and Diptiman Sen  for providing  useful references.
The support of Humboldt Foundation (Germany) is sincerely acknowledged.


\begin{thebibliography}{99}
\bibitem{ges}
F. D. M. Haldane,  Phys. Rev. Lett.  {\bf 67}, 937 (1991)
\bibitem{wuyu}
Y. S. Wu and Y. Yu,  Phys. Rev. Lett.  {\bf 75}, 890 (1995)
\bibitem{luttinger}
J. M. Luttinger, J. Math. Phys. (N. Y.) {\bf 4}, 1154 (1963)

F. D. M. Haldane, J. Phys. C  {\bf 14}, 2585 (1981)
\bibitem{wen} X. G. Wen, Int. J. Mod. Phys. B {\bf 6}, 1711 (1992)
\bibitem{fqhe}
D. S. Tsui, H.L. Stromer and A.C. Gossard, Phys. Rev. Lett. {\bf 48}, 1559
 (1982)
\bibitem{rabello}
S. J. B. Rabello,  Phys. Rev. Lett. {\bf 76}, 4007
 (1996)
\bibitem{jakiws}
U. Aglietti, L. Griguolo, R. Jackiw, S. Y. Pi and D. Seminara,
 Phys. Rev. Lett. {\bf 77}, 4406
 (1996)
\bibitem{gutkin} E. Gutkin, Annals of Phys. {\bf 176}, 22 (1987)
\bibitem{dipti} D. Sen, {\it Quantization of the derivative NLS} (unpublished)
\bibitem{delta} E. H. Lieb and W. Liniger, Phys. Rev. {\bf 130}, 1605 (1963)
\bibitem{snirman} A. G. Snirman, B. A. Malomed a E. B. Jacob, Phys. Rev.
 {\bf A 50}, 3453 (1994)
\bibitem{qnls} P. Kulish and E. K. Sklyanin, Lect. Notes. Phys. (ed. J.
Hietarinta et al, Springer, Berlin, 1982), {\bf 151}, 61.
\bibitem{qdnls} Anjan Kundu and B, Basu-Mallick, J. Math. Phys.{\bf 34}.
1252 (1993)
\bibitem{gutkin2} E. Gutkin, Ann. Inst. Henri-Poincar\'e, {\bf 2}, 67
(1985).
\bibitem{handbook} G. A. Korn and T. M. Korn, {\it Mathematical Handbook for
Scientists and Engineers} (McGraw-Hill, N.Y., 1961)
\bibitem{partha} P. Mitra,  Phys. Lett. {\bf B 345} , 473 (1995)
\bibitem{pla98} Anjan Kundu,  Phys. Lett. {\bf A 249} , 126 (1998)
\bibitem{ss-prl98}  H. J. Schulz, B. S. Shastry,  Phys. Rev. Lett. {\bf 80}
 , 1924 (1998)
\end{thebibliography}
 \end{document}